\documentstyle[12pt]{article}

\begin{document}
\vspace{-2.0cm}
\bigskip
\begin{center}
{\Large \bf
A Novel Approach to Noncommutativity in Planar Quantum Mechanics}
\end{center}
\vskip .8 true cm
\begin{center}
{\bf Rabin Banerjee}\footnote{rabin@bose.res.in},
 
\vskip 1.0 true cm 
S. N. Bose National Centre for Basic Sciences \\
JD Block, Sector III, Salt Lake City, Calcutta -700 098, India.
 
\end{center}
\bigskip
 
\centerline{\large \bf Abstract}
Noncommutative algebra in planar quantum mechanics  is shown to follow from 
't Hooft's recent analysis on dissipation and quantization. The 
noncommutativity in 
the coordinates or in the momenta of a charged particle in a magnetic field 
with an oscillator potential are shown as dual descriptions of the same
phenomenon. Finally, 
noncommutativity in a fluid dynamical model, analogous to the lowest
Landau level problem, is discussed.  
\section{Introduction}

Noncommutative spaces arise as brane configurations in string theory as
well as in the matrix model of M-theory \cite{review}. The appearance of this
noncommutativity is quite similar to a corresponding effect in the Landau
problem of a charged particle moving on a plane, subjected to a strong
perpendicular magnetic field. It is therefore desirable to investigate the
noncommutative effects in planar quantum mechanics 
to get a deeper understanding of this phenomenon in the 
context of string or matrix models. Some studies 
\cite{jl} along this direction have been performed. But these are done in a
way that either requires the occurence of noncommutativity (in the momenta)
due to the presence of a magnetic field, or just as a postulate 
(noncommutativity in  the coordinates).

   In this paper we would like to address the question of how and why such
noncommutative structures appear in planar quantum mechanics and provide
a systematic way of obtaining them. For this we adopt a recent approach 
advocated by 't Hooft \cite{hooft1, blasone} to study quantum mechanics. In
this approach the classical equations are written in a formulation using
quantum mechanical notation. However the quantum mechanical hamiltonian does
not have a lower bound, which is the price to be paid for expressing a 
classical system quantum mechanically. Constarints, motivated by information
loss or dissipation, are now imposed leading to a well defined bounded 
hamiltonian as well as to an apparent quantisation of the orbits.

  The point we make is that a modification of the hamiltonian brought 
about by  dissipation leads to a deformation in the symplectic structure.
The original Poisson brackets are changed to a new set of brackets. This
 is worked out in details  for the motion of  a charged particle in a 
plane subjected to a magnetic field(Landau problem) and a quadratic potential.
It naturally leads to a noncommutative algebra which is otherwise postulated.
Thus, apart from illuminating the various noncommutative structures, the 
present work also places 't Hooft's observations in a different setting.

 In section 2, we review the emergence of noncommutative coordinates 
for the motion of particles in the lowest 
Landau level. A strong magnetic field is necessary to effect the projection 
to this level. However, we show that it is possible to map
 the general Landau problem to a chiral oscillator model 
which is described by noncommuting variables. 
Thus the Landau problem for any magnetic field exhibits
 noncommutativity, albeit in a different set of coordinates.
Indeed the mapping of the Landau problem to the chiral oscillator has also
been used in the subsequent analysis.

The noncommutativity in the general problem of a charged particle in magnetic 
and electric fields, mentioned earlier, is discussed in section 3. This problem 
has recently generated considerable interest in both theoretical studies 
\cite{jl} and phenomenological applications \cite{me}. The second order 
equation of motion is expressed as a pair of first order equations. In this 
version, 't Hooft's analysis, a synopsis of which is given below, is applicable.
 Noncommutativity in the coordinates or in the momenta are shown to be dual 
descriptions, corresponding to distinct polarisations chosen for converting 
a second order system to a first order system. The issue is further clarified
 by interpreting the results as different superpositions of the noncommutativity
 in the chiral oscillator models of section 2.    

In section 4, a magnetohydrodynamical(MHD) field theory will be investigated 
from 't Hooft's formalism. A complete mapping with the lowest Landau level 
problem will be done, providing an alternative derivation of noncommutative 
MHD \cite{zg}. 

We present some concluding remarks in section 5.

We end this section by briefly mentioning  't Hooft's analysis. This is 
basically a hamiltonian formulation where first order equations
 $\dot{q}_i = \{q_i, H\} = f_i(q)$, decoupled from the conjugate momenta $p_i$, are considered.
The corresponding hamiltonian $H= \sum_{i} p_i f_i(q)$ obviously does not have
 a  lower bound. This is cured by expressing $H$ as, 
\begin{equation}
H = H_+ -H_-
\label{1}
\end{equation}
where   
\begin{equation} 
H_\pm = \frac{1}{4\rho} (\rho \pm H)^2 
\label{2}
\end{equation}
and $\rho$ is a positive definite function of $q_i$ satisfying
\begin{equation}
 \{\rho, H\} = 0
\label{3}
\end{equation}
so that ordering ambiguities (upon quantization) in (\ref{2}) are avoided. To 
get a lower bound for $H$, one imposes the constraint 
\begin{equation} 
H_-|\psi> = 0 
\label{4}
\end{equation} 
on the physical space. The constraint can be motivated by dissipation or
information loss \cite{hooft1}. In this paper however we shall not be
concerned with studying dissipative effects, but rather dwell on a completely
new manifestation that has not been discussed either in the original
papers \cite{hooft1} or in the subsequent applications \cite{blasone}.
The point is that a change from the original (unbounded) hamiltonian $H$
to the bounded positive (semi) definite hamiltonian $\rho$ leads to  
a modified algebra that can be obtained as follows,
\begin{equation} 
\dot{q}_i = \{q_i, \rho\} =  \{q_i, q_j\} \partial_j \rho(q)
\label{5}
\end{equation}
To reproduce the original set of equations of motion, obviously $q_i$ can
no longer be taken as commuting and must be calculated from,

\begin{equation}
\{q_i, q_j\} \partial_j \rho(q) =  f_i(q)
\label{6}
\end{equation}
leading to explicit noncommutative structures for the algebra of $q_i$.
Furthermore if the brackets $\Lambda_{ij} = \{q_i, q_j\}$ are nondegenerate, 
the corresponding lagrangian can also be evaluated,
\begin{equation}
L =  \frac{1}{2}q_i\Lambda^{ij}\dot{q}_j - \rho(q)
\label{7}
\end{equation}
where $\Lambda^{ij}$ is the inverse of $\Lambda_{ij}$,
\begin{equation} 
\Lambda^{ij} \Lambda_{jk} = \delta^i_k
\label{8}
\end{equation}
\section{The Landau  problem revisited}
The classical equations of motion for an electron of charge $-e$ moving in the 
$x_1-x_2$ plane under the influence of a constant perpendicular magnetic field 
$B$ are,
\begin{equation}
m\ddot{x}_i = -\frac{e}{c} B \epsilon_{ij}\dot{x}_j
\label{9}
\end{equation}
whose general solution in the complex notation $Z = x_1 + ix_2$, corresponds to
motion in a circle of radius R, 
\begin{equation} 
Z = R e^{i(\omega_c t + \delta)}
\label{9+1}
\end{equation}
Here $\delta$ is an arbitrary phase and the cyclotron frequency is, 
\begin{equation} 
\omega_c = \frac{eB}{mc}
\label{10}
\end{equation}
The radius of the orbit is related to the tangential velocity $v$ by $v= \omega_c R$. The centre of the orbit is chosen at the origin, but it could be 
displaced to  any other position $C (Z \rightarrow Z+C)$ without changing the 
physics. 

The above equations of motion follow from the Lagrangian,
\begin{equation}
L =  \frac{m}{2}\dot{x}^2_i - \frac{e}{2c} B  \epsilon_{ij}x_i\dot{x}_j  
\label{11}
\end{equation} 
The canonical momentum is given by, 
\begin{equation}
p_i = \frac{\partial L}{\partial \dot{x}_i} = m\dot{x}_i + \frac{e}{2c} B  \epsilon_{ij}{x}_j 
\label{12}
\end{equation} 
while the hamiltonian is 
\begin{equation}
H = \frac{{\pi}^2_i}{2m} = \frac{1}{2m}\left(p_i - \frac{e}{2c} B \epsilon_{ij}{x}_j \right)^2
\label{13}
\end{equation}
In analogy with the classical considerations it is possible to define a 
quantum orbit centre operator, 
\begin{equation}
C = Z + \frac{i\pi}{m\omega}_c; C_i = x_i +\frac{i\pi_i}{m\omega}_c
\label{14}
\end{equation} 
which is the source of the degeneracy in the Landau problem and satisfies the 
algebra,
\begin{equation}
\left[C_i, C_j\right] = i\frac{\hbar c}{eB} \epsilon_{ij}
\label{15}
\end{equation}

Let us now specialise to the lowest Landau level problem by enforcing a strong 
magnetic field. Heuristically, the classical analysis shows that the particles 
are projected to their respective orbit centres $(Z =C )$, where they are frozen.
The kinetic energy is completely quenched, but the algebra among the 
co-coordinates, given by (\ref{15}), is noncommutative. 
The above result can be understood by naively setting $m= 0$ in (\ref{11}) (quenching of the kinetic term) in which case the lagrangian is first order
 leading to the fundamental brackets, 
\begin{equation}\left[x_i, x_j\right] = i\frac{\hbar c}{eB} \epsilon_{ij}
\label{16}
\end{equation}
reproducing (\ref{15}).

A more rigorous edge to these arguments is provided by using the projection 
technique developed in \cite{sg}. The basis of states in the lowest Landau level
may be chosen as, 
\begin{equation}
\psi_n(Z,Z ^{\ast}) = \frac{Z^n}{\sqrt{{(2l^2)}^n n! \pi}}{\exp}\left(-\frac{ZZ^{\ast}}{4l^2}\right)
\label{17}
\end{equation}
where the magnetic length $l$ is defined by,
\begin{equation}
l^2 = \frac{\hbar c }{eB}
\label{18}
\end{equation}
The exponential factor, which is common to all eigenfunctions, is 
removed by formally defining a  Hilbert space of analytic functions\cite{sg}.
One now works with the polynomial factor only which, in contrast to the 
exponential factor, is analytic. In this scenario the projection of $Z$ 
is trivial $(Z\rightarrow Z)$ since it is already 
contained in the lowest Landau level. 
The projection of $Z^{\ast}$, on the other hand, leads 
to a differential operator \cite{sg}
$Z^{\ast} \rightarrow -2l^2\frac{\partial}{\partial Z}$.
The fundamental algebra, after this projection, thus yields, 
\begin{equation}
\left[x_1, x_2\right] =  \frac{1}{2i}\left[Z^{\ast}, Z\right] \rightarrow 
\frac{1}{2i}\left[-2l^2\frac{\partial}{\partial Z}, Z\right] = il^2
\label{19}
\end{equation}
showing the noncommutativity (\ref{16})
among the coordinates.

Having reviewed the noncommutativity in the lowest Landau level, let us
 provide a mapping of the original Landau problem which manifests this 
property for any magnetic field. Classically or quantum mechanically,
the Landau problem reduces to the motion of a one dimensional harmonic
 oscillator. Actually since the direction (clockwise or anticlockwise)of 
cyclotron motion of the charged particle is governed by the direction of the 
magnetic field(pointing up or down) one should regard it as the chiral form of
 the harmonic oscillator. The Lagrangian of a chiral oscillator is 
given by,
\begin{equation}
L_+ = \frac{1}{2}\epsilon_{ij} y_i \dot{y}_j - \frac{\omega}{2}y^2_i
\label{20}
\end{equation}
Eliminating any of the $y_i$'s in favor of the other yields a 1-d harmonic
oscillator in that variable with frequency $\omega$. The equation of motion is,
\begin{equation} 
\dot{y}_i = - \omega\epsilon_{ij} y_j 
\label{21}
\end{equation}
It is easy to read-off the brackets and the hamiltonian directly from (\ref{20})
\begin{equation} 
\{y_i, y_j\} = -\epsilon_{ij}
\label{22}
\end{equation} 
\begin{equation}
H = \frac{\omega}{2} y_i^2
\label{23}
\end{equation}
leading immediately to (\ref{21}) using the hamilton equations.

The chirality comes from the sign of the kinetic term in (\ref{20}).  Computing the
angular momentum, 
\begin{equation}
J = \epsilon_{ij}y_i p_j = \frac{H}{\omega} 
\label{24}
\end{equation}
where $p_i = \frac{\partial L}{\partial \dot{y}_i}$ is the conjugate momentum.
If the sign of the kinetic term is flipped so that 
\begin{equation}
L_- = -\frac{1}{2} \epsilon_{ij}y_i \dot{y}_j - \frac{\omega}{2} y_i^2
\label{25}
\end{equation}
we obtain $J = -\frac{H}{\omega}$. Note that $L_-$ also reduces to a 
1-d harmonic oscillator by the variable elimination process. Thus $L_\pm$ characterize a pair of chiral oscillators
rotating in an anticlockwise or clockwise direction. A suitable combination
 of such oscillators just leads to a two dimensional oscillator and has
 been used in the study of Zeeman effect and duality symmetry\cite{bg}.

The Landau problem is now mapped to the chiral oscillator by using the 
correspondence, 
\begin{equation}
\sqrt{m}\epsilon_{ij}\dot{x}_j \rightarrow \sqrt{\omega}y_i
\label{26}
\end{equation}
where $\omega$ is identified with the cyclotron frequency $\omega_c$. As a check we find that the Landau hamiltonian maps to the chiral oscillator hamiltonian,
\begin{equation} 
 \frac{m}{2}\dot{x}_i^2 \rightarrow \frac{1}{2}\omega y_i^2
\label{27}
\end{equation}
The equation of motion(\ref{9}) goes over to (\ref{21})  while the 
noncommutative 
algebra of $\pi_i$ in (\ref{13}) reproduces (\ref{22}).
Thus there is a complete mapping between the Landau model and the chiral 
oscillator. The coordinates in the latter satisfy a noncommutative algebra. 
The result is true for any value of the magnetic field $B$.
The limit of the strong magnetic field, as necessary in usual treatments
of noncommutativity in the lowest Landau level problem, is now irrelevant in
this new set of variables.

It may be mentioned that the above mapping also follows by recalling
 the equivalence \cite{dj} between the self dual model, 
\begin{equation}
{\cal L}_{SD} = \frac{1}{2}f_{\mu}f^{\mu} - \frac{1}{2m} \epsilon_{\mu \nu \lambda}f^{\mu}\partial^{\nu}f^{\lambda}
\label{28}
\end{equation}
and the Maxwell-Chern-Simons theory,
\begin{equation} 
{\cal L}_{MCS} = -\frac{1}{4}F_{\mu \nu}F^{\mu \nu} + \frac{m}{2}\epsilon_{\mu \nu \lambda} A^{\mu}\partial^{\nu}A^{\lambda}
\label{29}
\end{equation}
where the basic field $f_{\mu}$ gets identified with the dual field tensor $F_{\mu}$, 
\begin{equation}
f_{\mu} = F_{\mu} = \epsilon_{\mu \nu \lambda}\partial^{\nu}A^{\lambda} 
\label{30}
\end{equation}
If a projection from the field theory models to the planar quantum mechanical
 models is done by ignoring the spatial derivatives and setting the time
components $f_0, A_0$ to zero, the lagrangians (\ref{28}) and (\ref{29}) 
correspond respectively to the chiral oscillator and the Landau model. 
Then the map  (\ref{30}) is the analogue of (\ref{26}). Other possibilities and
consequences of such a projection have been discussed in \cite{bk1}.

This might be an appropriate place of explicitly working with the 't Hooft 
mechanism to realize the noncommutativity in the chiral oscillator, 
because equation (\ref{21}) only involves the coordinates. The 
corresponding hamiltonian is,
\begin{equation}
H = -(\omega \epsilon_{ij} y_j)p_i
\label{31}
\end{equation} 
which yields (\ref{21}) upon using $\dot{y}_i = \{y_i , H\}$. This hamiltonian is not bounded from below. Following 't Hooft's prescription discussed earlier, the physical
hamiltonian $\rho$ is constructed so that it commutes with $H$. This is given by
\begin{equation}
\rho = \frac{\omega}{2} y_i^2; \hskip 1.0cm \{\rho , H \} = 0
\label{32}
\end{equation}
To reproduce (\ref{21}) the symplectic structure gets modified to (\ref{22}) so 
that  
\begin{equation} 
\dot{y}_i = \{y_i, \rho \} = -\omega \epsilon_{ij} y_j
\label{33}
\end{equation}
The noncommutative bracket (\ref{22}) is thereby reproduced, together with the 
hamiltonian (\ref{23}). 

\section{Noncommutativity in the Landau problem with an oscillator potential}

The equation of motion for a charged particle moving in the $x_1 - x_2$ plane
under the influence of a magnetic field $B$ with an oscillator potential
is given by,\footnote{We have rationalised $e = c = 1$.} 
\begin{equation}
m\ddot{x}_i -B\epsilon_{ij} \dot{x}_j + kx_i = 0 
\label{34}
\end{equation}
Recall that the harmonic oscillator was recast in the form of a chiral
oscillator by doubling the degrees of freedom which manifested noncommutative
 properties. Thus the first step is to rewrite the second order system into a 
pair of first order equations by doubling the degrees of freedom. There is a 
certain amount of flexibility in doing this, but for the moment we confine to the following choice. 
\begin{equation} 
\dot{x}_i = \frac{1}{m}q_i; \hskip 1.5cm  \dot{q}_i = B\epsilon_{ij} \dot{x}_j - kx_i 
\label{35}
\end{equation}
Also $q_i$ satisfies an equation identical to (\ref{34}),
\begin{equation}  
m\ddot{q}_i -B\epsilon_{ij} \dot{q}_j + kq_i = 0
\label{36}
\end{equation}
The hamiltonian that yields the above set of  first order equations is,
\begin{equation}
H = \frac{1}{m}q_i\pi^x_i + \left( \frac{B}{m}\epsilon_{ij} {q}_j - kx_i \right)\pi^q_i
\label{37}
\end{equation} 
where $\left( x_i , \pi^x_i\right)$ and $\left( q_i , \pi^q_i \right)$ are
canonical pairs. This hamiltonian is not bounded from below. Following 't Hooft,
 a bounded hamiltonian follows upon dissipation. Explicitly this is found by 
constructing a positive definite $\rho$ that commutes with $H$:
\begin{equation}
\rho = \frac{1}{2}\left( \frac{q^2_i}{m} + kx^2_i\right); \hskip 1.5cm \{\rho , H \} = 0 
\label{38}
\end{equation}
In order to generate the equations (\ref{35}) from $\rho$ ($\dot{x}_i = \{x_i , \rho \}, \dot{q}_i = \{q_i , \rho \}$), it is obvious that the basic brackets 
are now given by,
\begin{equation} 
\{x_i, q_j \} = \delta_{ij}, \hskip 1.0cm \{x_i, x_j \} = 0, \hskip 1.0cm \{q_i, q_j \} = B\epsilon_{ij}
\label{39}
\end{equation} 
thereby revealing a noncommutative algebra. To make contact with known results,
one interprets the hamiltonian $\rho$ in (\ref{38}) as the usual harmonic oscillator 
hamiltonian with noncommutative momenta $(q_i)$ manifesting the presence of 
the magnetic field. It is straightforward to obtain the lagrangian since the 
matrix,
\begin{equation}
\Lambda_{ij} = \left[\{\Gamma_i, \Gamma_j\}\right] = \left( \begin{array}{cc}
                                               0 & \delta_{ij} \\
                                            -\delta_{ij}  & B\epsilon_{ij} 
                                          \end{array} \right)
\label{40}
\end{equation} 
of the coordinates $\Gamma = (x, q)$ possesses an inverse,
\begin{equation}
\Lambda^{ij} = \left( \begin{array}{cc} 
                         B\epsilon_{ij} & -\delta_{ij} \\
                         \delta_{ij}   & 0 
                          \end{array} \right)
\label{41}
\end{equation} 
The lagrangian is given by, 
\begin{equation}
L = \frac{1}{2} \Gamma_i \Lambda^{ij} \dot{\Gamma}_j - \rho(\Gamma )
  = q_i \dot{x}_i + \frac{B}{2} \epsilon_{ij} x_i \dot{x}_j - \frac{1}{2}
\left( \frac{q^2_i}{m} + kx^2_i \right) 
\label{42}
\end{equation}
Explicitly the equations of motion from the lagrangian reproduce (\ref{35}). 
Since $q_i$ is an auxiliary variable it can be eliminated by solving for it 
to yield, 
\begin{equation} 
L = \frac{m}{2} \dot{x}^2_i + \frac{B}{2}\epsilon_{ij}x_i\dot{x}_j - \frac{1}{2}kx_i^2 
\label{43}
\end{equation}
which is the standard lagrangian leading to (\ref{34}).

Next, we show how noncommutativity in the coordinates, instead of the momenta, 
arise. As remarked before there is a freedom in splitting (\ref{34}) into a
dual set of first order equations. Consider the following polarisation,
\begin{equation}
\dot{q}_i = -kx_i , \hskip 1.5cm \dot{x}_i = \frac{q_i}{m} + \frac{B}{m}\epsilon_{ij}x_j 
\label{44}
\end{equation} 
This set also yields the equation (\ref{34}) as well as (\ref{36}). The desired
 hamiltonian is, therefore, given by, 
\begin{equation}
H = (-kx_i)\pi^q_i + \left(\frac{q_i}{m} + \frac{B}{m}\epsilon_{ij}x_j \right)
\pi^x_i 
\label{45}
\end{equation}
The physical hamiltonian $\rho$ is once again given by (\ref{38}). However 
since the basic equations(\ref{44}) are different from (\ref{35}) the algebra (\ref{39}) also gets 
modified to, 
\begin{equation} 
\{x_i , q_j \} = \delta_{ij}, \{q_i , q_j \} = 0, \{x_i , x_j \} = \theta \epsilon_{ij}
\label{46}
\end{equation} 
where the noncommutativity parameter $\theta = \frac{B}{km}$. Comparing with
 (\ref{39}) shows that the noncommutativity has been shifted from the 
momenta to the 
coordinates.

As before, the lagrangian can be computed from an inverse Legendre transform.
The result is, 
\begin{equation}
L = q_i \dot{x}_i + \frac{\theta}{2}\epsilon_{ij}q_i\dot{q}_j - \frac{k}{2}x_i^2
 - \frac{q_i^2}{2m} 
\label{47-1}
\end{equation} 
leading to the set of equations (\ref{44}). Eliminating the auxiliary  variable
$x_i$ yields, 
\begin{equation}
L = \frac{\dot{q}^2_i}{2k} + \frac{\theta}{2}\epsilon_{ij} q_i\dot{q}_j - 
\frac{q^2_i}{2m}
\label{47}
\end{equation} 
from which equation (\ref{36}) follows directly. 

The above analysis shows that noncommutativity is a manifestation of 
the presence of the magnetic field. In the extended set of coordinates this 
noncommutativity can occur either in $x_i$ or in $q_i$. However it is also 
possible to interpret $(x_i, q_i)$ as phase space variables. The common 
structure of the hamiltonian $\rho$ shows that $x_i$ and $q_i$ are coordinates
and momenta, respectively. 
 Interpreted this way, noncommutative 
spatial   variables are just the dual of what happens for the momenta.
Intrinsically there is no fundamental difference between them. 
Depending on the polarisation it is possible to have noncommuting momenta or 
coordinates. The dual nature is further realised by observing that
while one polarisation yields the lagrangian in the $x$ variables, the other 
yields the lagrangian in the $q$ variables. The duality map 
 $m \leftrightarrow k^{-1}$ connects the lagrangians (\ref{43}) and 
(\ref{47}); i.e., small `$m$' in 
(\ref{43}) corresponds to large `$k$' in the latter. Now for  small `$m$',
(\ref{43}) reduces to a chiral oscillator for which 
$\{x_i, x_j\} = - B^{-1} \epsilon_{ij}$. In \cite{zg}, a canonical derivation 
of this result is provided by observing that the hamiltonian (\ref{38}) is
now meaningful only if $q_i = 0$. Imposing this as a constraint in (\ref{39}),
and then calculating the Dirac bracket, which, for any pair of 
variables $X_i, X_j$ subjected to the constraints $\Omega_l$, is defined as,
\begin{equation}
\{X_i, X_j\}_{DB}= \{X_i, X_j\}-\sum\{X_i, \Omega_l\}W^{lm}\{\Omega_m, X_j\}
\label{new}
\end{equation}
where $W^{lm}$ is the inverse of the bracket among the constraints
$(\{\Omega_l, \Omega_m\})$,one finds \cite{zg},
\begin{equation}
\{x_i, x_j\}_{DB} = -\sum \{x_i, q_l\} W^{lm} \{q_m , x_j\}= W^{ij}
\label{47+1}
\end{equation}  
where $W^{ij}  =-B^{-1} \epsilon_{ij} $. 
The noncommutative bracket is
 thus obtained as a Dirac bracket. 

In the dual version small `$m$' is replaced by large` $k$'. Then (\ref{38})
is meaningful for $x_i = 0$. Imposing this as a constraint in (\ref{46}), 
the Dirac bracket among $q_i$ is,
\begin{equation} 
\{q_i, q_j\}_{DB} = - \sum\{q_i, x_l\} \tilde{W}^{lm} \{x_m, q_j\} 
= \tilde{W}^{ij}
\label{47+2}
\end{equation}
where $\tilde{W}^{ij} = -\theta^{-1} \epsilon_{ij} $ is the inverse of 
$\{x_i, x_j\}$. This Dirac bracket is exactly the noncommutative bracket 
obtained direcly from  (\ref{47}) for large `$k$'. This completes the analysis 
of the dual structure. 

We now show that the origin of noncommutativity is contained in a similar 
feature occurring in the chiral oscillators considered in the previous section.
 Two such oscillators $(L_\pm)$ rotating in clockwise and anticlockwise 
directions, respectively, with different frequencies, exactly simulate the 
Landau problem in the presence of the oscillator potential. This was shown in
\cite{bk1}. The magnetic field gets identified with the difference in the 
frequencies of the oscillators.    
Naturally, if the fequency of the oscillators is identical, the doublet of
chiral oscillators just reduces to a two dimensional harmonic oscillator.
Consider, therefore the equations of motion for the chiral oscillators($L_\pm$)
(see \ref{20}) with frequencies $\omega_\pm$:
\begin{equation}
\dot{z}_i = \omega_+\epsilon_{ij}z_j, \hskip 1.5cm \dot{y}_i = -\omega_-\epsilon_{ij}y_j
\label{48}
\end{equation} 
satifying the algebra,
\begin{equation}
\{z_i, z_j \} = -\{y_i, y_j \} = \epsilon_{ij}; \hskip 1.5cm \{y_i, z_j \} = 0
\label{49}
\end{equation}
Expressing the above set of equations in terms of a new set of variables,
\begin{equation}
z_i + y_i = x_i
\label{50}
\end{equation} 
we find
\begin{equation} 
\dot{x}_i = \epsilon_{ij}(\omega_+  z_j - \omega_- y_j)
\label{51}
\end{equation} 
To reproduce the first part of (\ref{35}), put
\begin{equation}
 \epsilon_{ij}(\omega_+  z_j - \omega_- y_j) = \frac{1}{m}q_i
\label{52}
\end{equation}
Solving for $y_i$ and $z_i$ in terms of $x_i$ and $q_i$ yields,
\begin{equation}
y_i = \frac{1}{\omega_+ + \omega_-}(\omega_+ x_i + \frac{1}{m}\epsilon_{ij}q_j)\label{53}
\end{equation} 
\begin{equation}  
z_i = \frac{1}{\omega_+ + \omega_-}(\omega_- x_i - \frac{1}{m}\epsilon_{ij}q_j)
\label{54}
\end{equation} 
If we take the time derivative of the difference 
of the above two equations and exploit (\ref{48}), 
the second equation of (\ref{35}) is exactly reproduced 
with the following identification, 
\begin{equation}
k = m\omega_+  \omega_-, \hskip 1.5cm B = m(\omega_+ - \omega_-)
\label{55}
\end{equation} 
Expectedly the magnetic field is proportional to the difference of the 
frequencies. 

Knowing the algebra of $z_i$ and $y_i$, it is easy to get the algebra of 
$x_i$ and $q_i$, which is given by,
\begin{equation} 
\{x_i, x_j\} = 0 
\label{56}
\end{equation}
\begin{equation}
\{x_i, q_j \} =  m(\omega_+ + \omega_-) \delta_{ij}
\label{57}
\end{equation} 
\begin{equation} 
\{q_i, q_j\} = m^2(\omega_+^2 - \omega_-^2) \epsilon_{ij} =  m(\omega_+ + \omega_-) B \epsilon_{ij}
\label{58}
\end{equation} 
where the mapping (\ref{55}) has been used. Upto a trivial scaling this 
reproduces the algebra (\ref{39}). 

The hamiltonian of the composite system is obtained by adding the contributions 
from the two components and using (\ref{53}, \ref{54}),
\begin{equation} 
H = H_+ + H_- = \frac{\omega_+}{2} z_i^2 +  \frac{\omega_-}{2}y_i^2 = 
 \frac{1}{2m(\omega_+ + \omega_-)} \left(\frac{q_i^2}{m} + kx_i^2 \right)
\label{58n}
\end{equation}
which, modulo the scaling mentioned above, reproduces the 
desired expression (\ref{38}).

Proceeding similarly it is possible to reproduce the polarisation (\ref{44}) 
along with the associated algebra (\ref{46}). In this case define, 
\begin{equation}
z_i + y_i = q_i 
\label{58+1}
\end{equation}
and, 
\begin{equation}
\epsilon_{ij} (\omega_+ z_j - \omega_- y_j ) = -kx_i 
\label{58+2}
\end{equation}
so that the first equation in (\ref{44}) is obtained. The same steps adopted 
earlier lead to the second equation in (\ref{44}) with exactly the same 
normalization (\ref{55}).

The algebra of $q_i$ and $x_i$ follows from (\ref{58+1}) and (\ref{58+2}):
\begin{equation} 
\{q_i , q_j \} = 0; \hskip 1.0cm \{x_i , q_j \} = \frac{\omega_+ + \omega_-}{k} \delta_{ij}; \hskip 1.0cm \{x_i, x_j \} = \left(\frac{\omega_+ + \omega_-}{k}\right) \theta \epsilon_{ij}
\label{58+3}
\end{equation}
and a simple    scaling reproduces the desired algebra  (\ref{46}).
 
This section is concluded by providing a more general type of noncommutativity
involving both $x_i$ and $q_i$. Consider the pair of first order equations,
\begin{equation}
\dot{x}_i = \alpha \epsilon_{ij} x_j + \beta q_i 
\label{59}
\end{equation} 
\begin{equation} 
\dot{q}_i = \lambda \epsilon_{ij}q_j + \gamma x_i 
\label{60}
\end{equation}
which lead to the Landau type equations in both $x_i$ and $q_i$,
\begin{equation}
\ddot{r}_i = (\alpha + \lambda) \epsilon_{ij}\dot{r}_j + (\alpha \lambda + \beta \gamma)r_i; \hskip 1.5cm r_i = x_i, q_i 
\label{61}
\end{equation}
Following 't Hooft, a hamiltonian is constructed, 
\begin{equation}
H = (\alpha \epsilon_{ij}x_j + \beta q_i)\pi^x_i + (\lambda  \epsilon_{ij} q_j +  \gamma x_i) \pi^q_i 
\label{62}
\end{equation}
where $(x_i , \pi^x_i)$ and $(q_i , \pi^q_i)$ are canonical pairs. The 
equations of motion $\dot{r}_i = \{r_i, H \}$ just yields (\ref{59}, \ref{60}).
As usual, this $H$ is not bounded from below. A positive definite $\rho$, 
commuting with $H$, has to be obtained. A natural choice is, 
\begin{equation} 
\rho = \frac{1}{2}(ax_i^2 + bq_i^2); \hskip 1.5cm a,b \geq 0
\label{63}
\end{equation} 
where $a\beta + b\gamma = 0$ follows on demanding $\{\rho , H \} = 0$. To satisfy 
the condition among the parameters $\beta$ is chosen to be negative 
$(\beta = -k)$. Then the desired hamiltonian becomes, 
\begin{equation}
H \rightarrow  \rho = \frac{a}{2} (x_i^2 + \frac{k}{\gamma}q_i^2 )
\label{64}
\end{equation} 
The linearity of the original pair of first order equations show that they 
remain unchanged under a scaling $r_i \rightarrow \sqrt{\frac{\gamma}{a}} r_i$.
Moreover  to make contact with the familiar form of the hamiltonian we set 
$\gamma = m^{-1}$ where $m$ is a mass parameter. Then the hamiltonian reduces to 
\begin{equation}
\rho = \frac{1}{2} \left( \frac{x_i^2}{m} + kq_i^2 \right)
\label{65}
\end{equation} 
It is easy to check that the corresponding algebra is,
\begin{equation} 
\{q_i , x_j\} = \delta_{ij}; \hskip 1.0cm  \{x_i , x_j\} = \alpha \epsilon_{ij}; \hskip 1.0cm \{ q_i , q_j \} = \frac{\lambda}{k} \epsilon_{ij}
\label{66}
\end{equation} 
and the equations of motion (\ref{59}, \ref{60}) reduce to, 
\begin{equation} 
\dot{x}_i = \{x_i , \rho\} = \frac{\alpha}{m}\epsilon_{ij} x_j - kq_i
\label{67}
\end{equation}
\begin{equation} 
\dot{q}_i = \{q_i , \rho\} =\lambda \epsilon_{ij}q_j + \frac{1}{m}x_i
\label{68}
\end{equation}
Both $x_i, q_i$ satisfy a noncommuting algebra.

The Lagrangian follows from an inverse Legendre transform. The matrix 
constructed from the brackets is, 
\begin{equation}
\Gamma_{ij} = \left[\{r_i , r_j\} \right] = \left(\begin{array}{cc}
 \frac{\lambda}{k} \epsilon_{ij}  & \delta_{ij} \\
-\delta_{ij} & \alpha \epsilon_{ij} \end{array}
\right)
\label{69}
\end{equation}
whose inverse is given by
\begin{equation}
\Gamma^{ij} =  \frac{1}{k - \lambda \alpha}\left( \begin{array}{cc}
k \alpha \epsilon_{ij} & - k \delta_{ij} \\
 k \delta_{ij}  & \lambda \epsilon_{ij} 
\end{array}
\right)
\label{70}
\end{equation} 
Hence the lagrangian is given by,
\begin{equation}
L = \frac{1}{2} r_i \Gamma^{ij} \dot{r}_j - \rho
   = \frac{1}{2(k - \lambda \alpha)}\left[\lambda \epsilon_{ij} x_i \dot{x}_j 
+ \alpha k \epsilon_{ij} q_i \dot{q}_j + 2kx_i\dot{q}_i\right] - \frac{1}{2} 
\left[\frac{x_i^2}{m} + kq^2_i\right]
\label{71}
\end{equation}
Obviously $\frac{\lambda \alpha}{k} =1$ is a special point manifesting a 
degeneracy. This particular point has been discussed in details
elsewhere \cite{e}.
The equations of 
motion obtained from this lagrangian are compatible with (\ref{67}, \ref{68}).

The rotational symmetry of the hamiltonian $\rho$ suggests that the angular momentum 
operator $J$ has the conventional algebra with $x_i$ and $q_i$, 
\begin{equation} 
\{J, r_i\} = \epsilon_{ij}r_j; \hskip 1.5cm r_i = x_i, q_i 
\label{72}
\end{equation}
The structure of $J$ which yields these equations, from the basic brackets 
(\ref{66}), is given by, 
\begin{equation}
J = \frac{1}{2(k - \lambda \alpha)} \left( \lambda x_i^2 + \alpha k q_i^2 +
 2k\epsilon_{ij}q_ix_j \right)
\label{74}
\end{equation} 
Actually this expression for $J$ follows from (\ref{71}) by following a
standard canonical prescription,
\begin{equation}
J = \epsilon_{ij} (x_i \pi_j^x + q_i \pi^q_j )
\label{75}
\end{equation}
where,
\begin{equation}
\pi_i^x = \frac{\partial L}{\partial \dot{x}_i} = -  \frac{\lambda}{2(k - \lambda \alpha)} \epsilon_{ij} x_j 
\label{76}
\end{equation} 
\begin{equation}
\pi_i^q = \frac{\partial L}{\partial \dot{q}_i} = \frac{k}{2(k - \lambda \alpha)} (2x_i - \alpha  \epsilon_{ij} q_j)   
\label{77}
\end{equation}
are the respective conjugate momenta.

\section{Noncommutative field theory in the lowest Landau level}

In a recent paper \cite{zg} noncommutativity in a magnetohydrodynamical model 
was studied by taking the vanishing 
limit of a mass parameter, in analogy with the 
lowest Landau level problem. A similar study will be done here, based on
the ideas presented earlier. 

A field theory is obtained by introducing electron creation and annihilation 
operators in terms of which the density operator is $\eta(x) = \psi^{\dagger} 
\psi(x)$. Now the hamiltonian $h$ of a particle in the lowest Landau level is
just the projection of potential to that level \cite{gz}, 
\begin{equation} 
h = \int d^2 x V(x) \eta(x) 
\label{78}
\end{equation}
since the kinetic term, after projection, gets quenched. 

If a particle substructure is given to a fluid, then $\eta(x)$ is the density
 of the fluid satisfying the continuity equation, 
\begin{equation} 
\dot{\eta} = -\partial _i (\eta v_i) 
\label{79}
\end{equation}
The hamiltonian leading to this equation is,
\begin{equation} 
H = -\int d^2 x \partial_i (\eta v_i) \pi^{\eta}
\label{80}
\end{equation}
where $(\eta, \pi^{\eta})$ is a canonical pair.  As usual, this is not bounded
 from below.  The physical hamiltonian would be the operator commuting with 
$H$. In our case this is given by (\ref{78}) so that $\{h, H\} = 0$ yields,
\begin{equation}
\int d^2 x V(x) \partial_i (\eta v_i)(x) = 0 
\label{81}
\end{equation} 
Upto a normalization $N$, a solution for $v_i(x)$ is,
\begin{equation} 
v_i(x) = N \epsilon_{ij} \partial_j V(x)
\label{82}
\end{equation}
With this $v_i(x)$, the continuity equation (\ref{79}) is reproduced
 by bracketing with $h $ ($\dot{\eta} = \{\eta, h \}$) provided,
\begin{equation}
\{\eta(x), \eta(y)\} = -N \epsilon_{ij} \partial_i \eta \partial_j \delta (x-y)
\label{83}   
\end{equation} 
revealing a noncommutativity among the density operators. Exactly the same 
structure follows for the Landau problem if the explicit form for $\eta(x)$
and the noncommuting brackets among $x$ is used, with the identification,
\begin{equation}
N = \frac{c}{eB}
\label{84}
\end{equation} 
The bracket (\ref{83}) with the normalisation (\ref{84}) was obtained in
\cite{zg} by using the continuity equation and Euler equation in the massless
limit.

\section{Concluding remarks}
Let us now digress on the significance of our findings. The Landau model
lagrangians with an oscillator potential, 
expressed in their usual second order forms ({\ref{43}}) or 
({\ref{47}}) obviously satisfy
a conventional commutative algebra. In the standard treatments 
\cite{jl} one introduces
a noncommutative plane {\it by hand} and subsequently discusses the model on
that plane. Here we show that, by transforming the second order system to a
first order one, by introducing an additional variable, the noncomutative plane
is naturally induced. Indeed, by following 't Hooft's analysis done in a 
different context, we have provided a systematic way of obtaining the 
various noncommutative strutures on the plane. In this connection we might
recall that a ``natural explanation" for the occurrence of noncommutativity in
the Landau problem is attributed to the presence of the magnetic field. 
However, this effect is only found for the lowest Landau level in which case
the original second order system is effectively replaced by a first order one.
This is the origin of the change in the symplectic structure. Something 
similar happens here where a passage from the second order to the first 
order system has been effected. Our conclusions are further confirmed by
the fact that all the noncommuting structures were also obtained as a 
superposition of two chiral oscillators, which are essentially first order
systems, moving in opposite (clockwise and anticlockwise) directions.
Since the difference in the frequencies of the two rotations is actually
proportional to the magnetic field, the connection of this approach with
the one where noncommutaivity is just introduced 
through the presence of a magnetic field gets
established.

A consequence of our analysis has been that the manifestation of the 
noncommutativity in the coordinates or in the momenta is shown as a dual
aspect of the same phenomenon. Indeed we established a direct connection
between the noncommutativity parameter $B$ (the magnetic field) in the
momentum algebra with the corresponding parameter $\theta$ in the coordinate
algebra through the relation $\theta=\frac{B}{km}$.  
This is again reminiscent of an analogous connection found in the discussions
of noncommutativity in the context of open string quantisation \cite{review}.
Furthermore, the critical point (where the density
of states becomes infinte) found in the literature \cite{e} in the analysis 
comprising noncommutativity in both coordinates and momenta, was also obtained
here. At this point it was shown that the symplectic matrix does not have an
inverse so that the transition from the hamiltonian to the lagrangian by an
inverse Legendre transform was not feasible.

We conclude this paper by briefly suggesting the restrictions on the Landau
orbits brought about by the imposition of the constraints that eventually 
led to a bounded hamiltonian from the unbounded one. Thus if the period of
the cyclotron orbit in a chiral oscillator
be a function $T(\rho)$ of $\rho$, where the constraint leads to the physical 
hamiltonian $\rho = H$, then \cite{hooft1}
\begin{equation}
e^{-iHT}|\psi > = |\psi >
\label{85}
\end{equation}
so that, 
\begin{equation}\rho T(\rho) = 2 \pi n, \hskip 1.5cm n \in {\cal Z}
\label{86}
\end{equation}
This means that the orbits in the Landau problem with an oscillator potential
 comprise two pieces (see \ref{48}); clockwise orbits satisfying the 
periodicity condition, \begin{equation}
\frac{\omega_+}{2} z_i^2 T(z) = 2 \pi n, \hskip 1.5cm n \in {\cal Z}
\label{87}
\end{equation} 
and anticlockwise orbits obeying, 
\begin{equation}
\frac{\omega_-}{2} y_i^2 T(y) =  2 \pi m , \hskip 1.5cm m \in {\cal Z} 
\label{88}
\end{equation}
As a final remark we mention some recent works \cite{f} which reveal the 
continuing interest in quantum mechanics on the noncommuting plane. 

{\large \bf \noindent Acknowledgement} \\
I thank Gerard 't Hooft for a useful correspondence.

\end{document}